\documentclass{article}

\usepackage{bm}
\usepackage{latexsym}
\usepackage{dcolumn}
\usepackage{amsmath,amsfonts,amssymb}
\usepackage{graphicx}
\usepackage[toc,page]{appendix}

\def\eq#1{{Eq.~(\ref{#1})}}

\title{Thermality and Heat Content of horizons from infinitesimal coordinate transformations }
  \author{Bibhas Ranjan Majhi and T. Padmanabhan\\
  IUCAA, Pune University Campus,\\
  Ganeshkhind, Pune- 411 007.\\
  {\small {email: bibhas@iucaa.ernet.in};  {email: nabhan@iucaa.ernet.in}}
  }

\begin{document}

\maketitle

\begin{abstract}
Thermal properties of a static horizon, (like the entropy $S$, heat content $TS$ etc.) can be obtained either from the surface term of the Einstein-Hilbert action or by evaluating the Noether charge, corresponding to the diffeomorphisms generated by the timelike Killing vector field.  We show that, for a wide class of geometries, the same results can be obtained using the vector field which produces an \textit{infinitesimal} coordinate transformation between two physically relevant reference frames, viz. the freely falling frame near the horizon and the static, accelerated, frame. 
In particular, the \textit{infinitesimal} coordinate transformation from inertial coordinates to uniformly accelerated frame can be used to obtain the heat content and entropy of the Rindler horizon. This result offers insight into understanding  the observer dependent degrees of freedom which contribute to the entropy of null surfaces.
\end{abstract}



\section{\label{intro}Introduction}

Classical general relativity is built on the principle of general covariance which denies special status to any particular class of coordinate systems or associated observers. Phenomena which are specifically coordinate dependent are usually treated with suspicion by classical relativists. The situation, however, is different when one brings in quantum effects through the study of quantum field theory in non-trivial background spacetimes. The pioneering work by Davies and Unruh \cite{daviesunruh} showed that an accelerating observer will attribute a non-zero temperature to the quantum state, interpreted as a zero-temperature vacuum state by the inertial observers. Similarly the thermal phenomenon associated with a black hole will also depend on the observer; while a stationary observer  outside the event horizon will attribute  an entropy and temperature  to the event horizon, another observer freely falling through the event horizon will perceive the physics quite differently. It appears inevitable that these results require one to attribute an observer dependence to \textit{all} thermal phenomena in nature. (For more details on this point of view, see e.g.,\cite{TPreviews, tpstructure}).

The two physical contexts mentioned above --- the black hole horizon and the Rindler horizon --- are, in fact,  related to each other  in a fairly simple manner. The metric near the event horizon of a massive ($M\gg M_{Planck}$) black hole can be approximated by the metric of the accelerated (Rindler) frame in flat spacetime. The thermal phenomena perceived by observers close to the event horizon can then be mapped to those perceived by Rindler observers in flat spacetime and one obtains the correct expression for the temperature by this mapping. In this correspondence, the freely falling observers through the event horizon will be analogous to inertial observers in flat spacetime while the stationary observers outside the event horizon will be similar to uniformly accelerated observers in flat spacetime. Given the fact that stationary observers in black hole spacetime attribute an entropy density (entropy  per unit area) of 1/4 in Planck units to the event horizon, it seems inevitable that accelerated observers in flat spacetime will also attribute an entropy density of 1/4 to the Rindler horizon. The quasi-local nature of physics implies that any null surface which acts as a local Rindler horizon to a class of observers should have an observer dependent  entropy density of 1/4 vis-a-vis this class of observers.

While the above results do not seem to lead to any clear-cut paradoxes (like violation of conservation laws or perpetual motion)  they do appear  somewhat perplexing and counter-intuitive. In non-gravitational physics, we are accustomed to thinking of entropy as an observer-independent quantity being related to certain unobserved degrees of freedom. It is, therefore tempting to think of the entropy of a black hole as being absolute and is observer independent and related to some specific degrees of freedom. This, however, cannot be true in view of the argument given above, if the quasi-local nature of physics has to be maintained. The fact that an observer freely falling through the black hole event horizon has access to more spacetime region than the observer eternally confined to the outside of the black hole  shows how such an observer dependence can indeed arise in this context. While this is intuitively understandable, it would be nice to see how this effect arises in the standard procedures we adopt for the computation of the entropy of the horizons. We will illustrate one class of such examples in this paper.

A possible mechanism behind the origin of observer dependent gravitational degrees of freedom was suggested in Ref.\cite{bibhas-tp} based on the approach pioneered by Carlip [\cite{Carlip:1999cy}; also see \cite{Solodukhin:1998tc}]. The standard description of gravity is invariant under the set $\mathcal{C}$ of all possible diffeomorphisms. These diffeomorphisms allow us, in principle, to remove all the gauge degrees of freedom retaining only the diffeomorphism invariant physical degrees of freedom. When we limit our attention to a set of observers who perceive a null surface as a horizon, the physical context gets restricted to a situation in which only those diffeomorphisms in the set $\mathcal{C}'$, which retain the horizon structure, are allowed. While it is a difficult (and unsolved) problem to precisely quantify what this restricted class $\mathcal{C}'$ is, it seems reasonable to assume that $\mathcal{C}'$ is a proper subset of  $\mathcal{C}$. This, in turn, implies that using the diffeomorphisms in $\mathcal{C}'$,  we cannot remove all the redundant gravitational degrees of freedom which we could have originally removed using the full set $\mathcal{C}$. In other words, certain degrees of freedom which would have been treated as pure gauge (when the theory was invariant under $\mathcal{C}$) now gets `upgraded' to physical degrees of freedom (when the theory is invariant under 
$\mathcal{C}'$). The entropy attributed to the null surface, by the observers who perceive it as a horizon, could arise from these degrees of freedom. (See \cite{D'Hoker:2010hr} for related ideas in a different context.) 

The above description is purely intuitive and mathematically ill-defined. 
To make progress towards a somewhat concrete realization of the above ideas and test their conceptual veracity, we will investigate the following issue: In a flat spacetime, the inertial observers do not attribute any thermal properties to any null surface.  Consider now an \textit{infinitesimal} coordinate transformation $x^a\to x^a + q^a(x)$ from the inertial coordinate system to the Rindler coordinate system. We then ask: Is it  possible to obtain the thermal properties, associated with a null surface in flat spacetime, in terms of the vector field $q^a$? We will show that this is indeed possible and --- in this sense --- one can consider the infinitesimal transformations described by the vector field $q^a$ as having upgraded some of the gauge degrees of freedom to physical degrees of freedom. We stress that $q^a(x)$ is \textit{not} the usual Killing vector field associated with the Rindler time translation; in fact, \textit{we are not aware of  any previous discussion in the literature} of this particular vector field. The same ideas work in a much wider class of spacetimes with horizons, when we consider the infinitesimal coordinate transformations between the freely falling frame near the horizon and the frame of the static observers (like, for e.g., in terms of the infinitesimal coordinate transformation from Kruskal to Schwarzschild coordinates) and even in a more general context. 

The paper is organized as follows: In Section \ref{heat} we review and rephrase some of the standard known results to motivate and show how the surface Hamiltonian, $H_{\textrm{sur}}\equiv TS$ (which represents the heat containt of the horizon), is related to the surface action and the Noether charge. Next, we introduce the infinitesimal coordinate transformation from inertial to Rindler frame and the vector field corresponding to this transformation. In Section \ref{thermality}, both the surface term of the action and the Noether charge are evaluated in terms of this vector field 
and shown to lead to identical expression. We explain how this arises and provide a discussion of related issues. Section \ref{generalised} generalizes our findings for a general static, spherically symmetric metric and describes how a more general spacetime can be handled by this approach.  In the final section we discuss our results.

\section{\label{heat}Heat content and thermality  of  horizons}

We begin by  briefly reviewing some of the standard results rephrasing them in a useful manner. A convenient physical quantity which captures the thermality of a \textit{any} system is the difference between the energy $E$ and the free energy $F=E-TS$, viz. $TS$. In the case of a spacetime horizon, as we will see,  this  heat content can be related \cite{tpstructure,tpnoeeng} to the surface Hamiltonian 
\begin{equation}
H_{\rm sur} \equiv TS   = \frac{\kappa}{2\pi} \ S    = \frac{\kappa A_\perp}{8\pi}              
\end{equation}  
where $S=(A_\perp/4)$ is the entropy of the horizon with $A_\perp$ being its area, $T=(\kappa/2\pi)$ is the temperature with $\kappa $ being the relevant acceleration (Rindler acceleration or the surface gravity). In the case of non-compact horizons we will interpret $(H_{\rm sur}/A_\perp)$ as $Ts$ where $s$ is the entropy density, viz. entropy per unit area.  The acceleration $\kappa$ can be defined through the relation $\xi^b\nabla_b\xi^a = \kappa \xi^a$ in the case of Killing horizons associated with a time-like Killing vector field $\xi^a$, which becomes null on the horizon surface. Defined in this manner, the numerical value of $\kappa$ changes if we rescale $\xi^a$. This ambiguity is avoided by either defining the normalization of $\xi^a$ through some physical consideration or by dealing with quantities which are invariant under the rescaling. 

The heat content $H_{\rm sur}$ can be computed in two equivalent ways, both of which capture its physical meaning \cite{tpstructure,tpnoeeng}. The first method is to define $H_{\rm sur}$ as the Hamiltonian associated with the surface term of the Einstein-Hilbert action  through $H_{\rm sur} = -(\partial \mathcal{A}_{\rm sur}/\partial t)$. As is well-known, the Einstein-Hilbert Lagrangian can be written as a sum of a bulk term ($L_{bulk}$, which is quadratic in $\Gamma$) and a surface term $L_{sur}$ with  
\begin{equation}
 L_{\rm sur}=\frac{1}{16\pi}\partial_c(\sqrt{-g}V^c); \qquad V^c \equiv -\frac{1}{g} \partial_b(gg^{bc})
\end{equation}  
(see e.g., eq (6.15) of \cite{gravitation}). The surface action $\mathcal{A}_{\rm sur}$ is defined as the integral over $L_{sur}$. 
If the near horizon metric is approximated as a Rindler metric (with $-g_{00}=1/g_{xx}=N^2=2\kappa x$ to evaluate these expressions on $N=$const surface, we will get:
\begin{equation}
\mathcal{A}_{\rm sur}= \frac{1}{16\pi}\int_{x}dt d^2x_\perp V^x
=\pm t\left(\frac{\kappa A_\perp}{8\pi}\right)
\label{surfaceH}
\end{equation} 
where $A_\perp$ is the transverse area. (The sign depends on the convention chosen for the outward normal or whether the contribution of the integral is taken at the inner or outer boundaries; see e.g., the discussion in \cite{surH}. We will choose the negative sign in \eq{surfaceH}.)   More generally, a static, near-horizon, geometry can be described by the metric \cite{matt1,tpdawoodgentds}
\begin{equation}
ds^2=-N^2dt^2+dl^2+\sigma_{AB}dx^Adx^B;
\end{equation}
where $N$ and $\sigma_{AB}$ have the near-horizon behaviour of the form
\begin{equation}
 N=\kappa l +O(l^3); \quad\sigma_{AB}=\mu_{AB}(x^A)+O(l^2)
 \label{nofl}
\end{equation}  
with $l=0$ surface being the horizon. The integrals in \eq{surfaceH} again leads to the same result.
Near the static horizon, the integrand is independent of $t$ and transverse coordinates and hence $\mathcal{A}_{\rm sur}$ must be proportional to $tA_\perp$. The proportionality constant turns out to be $\kappa/8\pi$. 

With future applications in mind, we stress that the $A_{sur}$ used in \eq{surfaceH} is \textit{not} the full surface action but \textit{only} the contribution of it on the horizon.
This is relevant when $L_{sur}\propto \partial_c(\sqrt{-g}V^c)$ vanishes because $\sqrt{h}n_cV^c=$ constant. If we define the action as the integral of $L_{sur}$ over a spacetime domain $\mathcal{V}$, the action can be expressed as the surface integral over the boundary $\partial \mathcal{V}$ of this domain and this will also vanish.
But the contribution of the surface integral from \textit{part} of the boundary surface (like the horizon, say) will be given by the integral of $\sqrt{h}n_cV^c$ over that part of the surface and need not vanish. It is this contribution, computed over the horizon, which we have defined as $A_{sur}$ and this can be nonzero even when $L_{sur}\propto \partial_c(\sqrt{-g}V^c)=0$ provided $V^c$ itself is not zero. We will have occasion to consider such a case later on.

Given $A_{sur}$, the surface Hamiltonian is  defined in a standard manner as:
\begin{equation}
 H_{sur}\equiv -\frac{\partial\mathcal{A}_{sur} }{\partial t}
 =\frac{1}{16\pi}\int_{x} d^2x_\perp V^x
=\left(\frac{\kappa A_\perp}{8\pi}\right)=TS
\label{horsurfham}
\end{equation} 
with suitable choice of sign.
The $H_{sur}$ is the difference between the energy $E$ and the free energy $F=E-TS$ of a finite temperature system and measures the heat content of the horizon.

The second method \cite{tpstructure,tpnoeeng} to compute the heat content is by using the Noether potential. In any gravitational theory which is invariant under the diffeomorphism $x^a \to x^a + q^a(x)$, one can define a conserved Noether current $J^a$ (which depends on $q^a$) related to an anti-symmetric Noether potential $J^{ab}$ through $J^a  = \nabla_b J^{ab}$. In Einstein's theory one can choose the Noether potential to be:
\begin{equation}
 J_{ab} = \frac{1}{16\pi} \left( \nabla_a q_b - \nabla_b q_a \right) 
 = \frac{1}{16\pi} \left( \partial_a q_b - \partial_b q_a \right) 
\end{equation} 
The surface Hamiltonian can then be defined as the surface integral over a co-dimension-2 cross-section of the null surface:
\begin{equation}
H_{\rm sur} = TS \equiv    \int d^{D-2}\Sigma_{ab}\; J^{ab}
\label{noetherint}
\end{equation} 
where the Noether potential is computed for   $q^a = \xi^a$  with  $\xi^a$ being   the local Killing vector corresponding to time translation symmetry of the local Rindler frame near the horizon. (The connection between Noether charge and energy has been previously noted in, e.g., Refs. \cite{hayward})
Note that both $J_{ab}$ and $T$ rescale in the same manner, if $\xi^a$ is rescaled, leaving the entropy invariant. 

While the two ways of computing $H_{\rm sur}$ described above --- from the surface term of the Einstein-Hilbert action in \eq{horsurfham} or from the Noether potential  related to the time-like Killing vector in  \eq{noetherint} --- lead to the same expression for $H_{\rm sur}$, there is a subtle difference between the two results. The surface term of Einstein-Hilbert action defined through $L_{sur}$ is not generally covariant. In flat spacetime, $A_{sur}$ will vanish in inertial coordinates while it can be non-zero in the non-inertial coordinates. To obtain a non-zero value for $H_{\rm sur}$ from $\mathcal{A}_{\rm sur}$ using \eq{horsurfham} we need to explicitly work with the noninertial, Rindler, coordinates in flat spacetime. More generally, in a curved spacetime, the expression for $\mathcal{A}_{\rm sur}$ can depend on the coordinate system which is used.
The expression for $H_{\rm sur}$ in terms of the Noether potential in \eq{noetherint}, on the other hand, is manifestly covariant and will lead to the same result in all coordinate systems. While this may be surprising at first sight, it should be noted that --- while the value of the integral in \eq{noetherint} will be the same in all coordinate systems --- the physical interpretation of $\xi^a$ used to define the Noether potential is different in different coordinate systems. In flat spacetime,  $\xi^a$ will be thought of as generator of time translation in Rindler frame but it is (proportional to) the Lorentz boost generator in
the inertial frame.

It is worth mentioning at this stage that there is another  (York-Gibbons-Hawking, \cite{ygh}) surface term which is often discussed in the gravitational context and is given by
\begin{equation}
 \mathcal{A}_{\rm YGH} \equiv \pm\frac{1}{16\pi} \int_{\mathcal{S}} d^3 x \, \sqrt{h} \, (2K)
\end{equation} 
where $K$ is the extrinsic curvature of the surface $\mathcal{S}$ and $h_{ab}$ is the metric induced on this surface. (The overall sign depends on whether the surface is spacelike or timelike.) This term is usually added to the Einstein-Hilbert action to cancel the variations of the metric normal to the boundary. Contrary to the popular myths: (i) The action $\mathcal{A}_{\rm YGH}$ does \textit{not} cancel the surface term $\mathcal{A}_{\rm sur}$ in general. 
(ii) Even their \textit{variations} do not cancel each other for \textit{arbitrary} variations of the metric but they do cancel each other if the metric is held fixed on the boundary [see e.g., pages 249-250 of Ref.~\cite{gravitation}]. This property is shared by an infinite set of surface functionals \cite{charap}  and the choice of $\mathcal{A}_{\rm YGH}$ is merely a convenient one to make the variational principle in gravity well defined. In this sense, $\mathcal{A}_{\rm sur}$  has a better standing as a well defined surface term arising naturally from the Einstein-Hilbert action itself than  $\mathcal{A}_{\rm YGH}$. Nevertheless, given the popularity enjoyed by $\mathcal{A}_{\rm YGH}$ in the literature, we will briefly mention the corresponding results for this action. 

It is easy to show that (see, for e.g., Ex. (6.3) of Ref.~\cite{gravitation}), in general, the integrands of the surface integral in  $\mathcal{A}_{\rm sur}$ and $\mathcal{A}_{\rm YGH}$ are related as follows: 
\begin{equation}
V^a n_a =   2K + 2 h^{ab}\partial_b n_a - n^m h^{ns} \partial_n g_{sm}
\end{equation} 
 The second and third terms on the right hand side will not vanish for an arbitrary surface and coordinates  because of which $n_c V^c$ is obviously not a scalar even though $2K$ is. However, if the coordinate system is chosen such that the surface corresponds to, say, $x^1 =$ constant with the metric being block diagonal with respect to the $x^1$ coordinate (i.e., $g_{1k} =0$ for $k\ne 1$) then it is easy to show that  $V^a n_a =  2K$ (see, for e.g., the Appendix of Ref.~\cite{TPPR}). Because most of the metrics we deal with (and certainly the Rindler metric which is valid close to the horizon) belongs to this class, we can use $2K$ and $V^c n_c$ interchangeably for our purpose. This implies that, in the Rindler frame, we should get the same answer for $\mathcal{A}_{\rm YGH}$ and $\mathcal{A}_{\rm sur}$ when evaluated on a stretched horizon corresponding to $x=\epsilon$ with the limit $\epsilon \to 0$ taken in the end. This is indeed true since in the Rindler frame $K\sqrt{h}=\kappa$=constant, independent of $\epsilon$, leading to the result: 
\begin{equation}
 \mathcal{A}_{\rm YGH} = - t \left(\frac{\kappa A_\perp}{8\pi}\right)
 \label{yghrindler}
\end{equation} 
So, in the Rindler frame, we can also write the result in \eq{horsurfham} as 
\begin{equation}
 H_{\rm sur} = - \xi^a \nabla_a \mathcal{A}_{\rm YGH} =  \xi^a \nabla_a \left[\frac{1}{8\pi} \int_{\mathcal{S}} d^3 x \, \sqrt{h} \, K\right]
\label{gencovH}
\end{equation} 
This provides a generally covariant expression for the heat content as long as the expression on the right hand side is well defined. If we repeat the calculation in the \textit{inertial} coordinates, it turns out that we obtain the same result provided: (a) We perform the differentiation in \eq{gencovH} \textit{before} taking the $\epsilon \to 0$ limit. 
(b) We choose $\xi^a$ with components $\xi^a = \kappa(X,T,\mathbf{0})$ in the inertial frame. This is the standard Killing vector for generation of the Lorentz boosts scaled up by $\kappa$. The  inertial frame has no intrinsic $\kappa$ in it and we can obtain the same result for $H_{\rm sur}$ in both frames only through this scaling. We will come back to this issue a little later. 

 Given the fact that $\mathcal{A}_{\rm sur}=0$ in the inertial coordinates (because $\Gamma$s vanish), while   $\mathcal{A}_{\rm sur} \ne 0$ and is given by \eq{surfaceH} in the Rindler coordinates, 
 one might wonder what we get for $\mathcal{A}_{\rm YGH}$ if evaluated in inertial and Rindler frames.
As we said before, we get $\mathcal{A}_{\rm YGH} = \mathcal{A}_{\rm sur}$ if we evaluate both in the Rindler frame. If we compute 
 $\mathcal{A}_{\rm YGH}$ in the inertial coordinates taking the surface $\mathcal{S} $ to be given by $X^2 - T^2 = \epsilon^2$ then we get the result
\begin{equation}
 \mathcal{A}_{\rm YGH} = - \frac{A_\perp}{8\pi} \, {\rm tanh}^{-1} \left( \frac{T}{X}\right)\Bigg|_{\mathcal{S}}
 \label{yghmink}
\end{equation} 
which, of course, matches with the result in \eq{yghrindler} for non-zero $\epsilon$ because ${\rm tanh}^{-1}(T/X)=\kappa t$ ensuring general covariance. But
a naive limit of $\epsilon \to 0$ corresponding to $(T/X) \to \pm 1$ leads to a divergence in \eq{yghmink} so that this expression is not well defined when we take the null surface limit. (One can do slightly better  by working in Euclidean coordinates when one would have got $\tan^{-1}( T_E/ X)$ instead of the tanh$^{-1}(T/X)$. The null surface limit now corresponds to $T_E \to 0, X\to 0$ and the results depend on the order in which these limits are taken. A choice of taking $T_E\to 0$ first with $X$ finite will give  $\mathcal{A}_{\rm YGH} =0$ which will match with $\mathcal{A}_{\rm sur}=0$ in inertial coordinates but not with \eq{yghrindler}.)
Thus  we find that $\mathcal{A}_{\rm YGH}$ is not well defined in the null surface limit when evaluated in the inertial coordinates even though the heat content, obtained after the differentiation is carried out in \eq{gencovH}, remains well defined
in both coordinates. Because of these issues and the comments made earlier about the ad-hoc nature of $\mathcal{A}_{\rm YGH}$, we will not use it further in our discussion.

We will next consider the corresponding aspects in the case of $H_{\rm sur}$ evaluated  from \eq{noetherint} 
by  explicitly computing it from \eq{noetherint} in flat spacetime in both Rindler and inertial coordinates. In the local Rindler frame with $-g_{00}=1/g_{xx}=N^2=2\kappa x$
if we take 
\begin{equation}
\xi^a = \delta^a_0 , \quad d^2\Sigma^{ab} = \frac{1}{2} (n^a m^b - n^b m^a)\sqrt{\sigma}\, d^2x_\perp,                                                                                               \end{equation} 
where $n^a = (2\kappa x)^{1/2} \, \delta^a_1$, $m^a = - (2\kappa x)^{-1/2}\delta^a_0$, then a straightforward calculation gives 
\begin{equation}
 H_{\rm sur} = \int_{\mathcal{H}} d^2\Sigma_{ab}\, J^{ab} = \frac{1}{16\pi} \int d^2x_\perp\, \sqrt{\sigma} \, (2\kappa)
= \frac{\kappa A_\perp}{8\pi}
\end{equation} 

The above calculation was done in the Rindler coordinates which has the parameter $\kappa$ in the metric.  But since the integral defining $H_{\rm sur}$ is generally covariant we could  have evaluated the same in the inertial coordinates as well.
In the inertial coordinates ($T,X, \mathbf{X}_\perp$), the components of $\xi^a$ will be $\kappa (X, T, \mathbf{0})$ which represents the Lorentz boost vector with \textit{an arbitrary scaling factor} $\kappa$.  This vector becomes null on $X = T$ and hence $\xi^b \nabla_b \, \xi^a$ must be proportional to $\xi^a$ on this surface; that is, $\xi^b \nabla_b \, \xi^a = \lambda \xi^a$ on $X=T$. An elementary calculation shows that $\lambda =\kappa$ which can be thought of as the non-affinity parameter  associated with the generators of the null surface $X=T$. (Obviously there is no thermal interpretation for $\kappa$ in the inertial coordinates.) The relevant non-zero component of the Noether potential in the inertial frame is $J_{TX} = 2\kappa$ while the measure for transverse integration is just $d\Sigma^{TX} = \frac{1}{2}\sqrt{\sigma}d^2 x_\perp$. The integral in \eq{noetherint} now becomes 
\begin{equation}
 H_{\rm sur} = \frac{1}{16\pi}\, \int_{\mathcal{H}}d^2x_\perp\,\sqrt{\sigma}\, (2\kappa) 
=  \frac{\kappa A_\perp}{8\pi}
\end{equation} 
Thus we get the same result, as we should, when we use the same vector $\xi^a$ in either coordinate system.
If we had known that non-affinity parameter $\kappa$ has the interpretation as horizon temperature in the Rindler frame, we could have done the entire computation in the inertial coordinate itself. (In Section \ref{thermality} we will describe a similar procedure but with a \textit{different} vector field.)

There is a related issue, regarding the dimensions of $\xi^a$, which is worth noting. If we think of the coordinates $x^a$ having the dimensions of length, then it is natural to assume that the deformation vector $q^a$ appearing in the diffeomorphism $x^a \to x^a + q^a$ also has the dimensions of length. But it is conventional to take the time-like Killing vector $\xi^a$ in static spacetimes to be {\it dimensionless} with $\xi^a = \delta^a_0$ as we did in the above calculation. In that case, one cannot take $q^a = \xi^a$ unless we change the dimensions of either one of them. In the inertial coordinates it is more natural to work with $\chi^a = 
\xi^a/\kappa = (X, T, \mathbf{0})$ (which is the usual Killing vector corresponding to Lorentz boosts with the dimension of length). This is equivalent to setting $\kappa =1$ and the non-affinity parameter for the null congruence generating the $X=T$ surface becomes unity;
that is, $\xi^b \nabla_b \, \xi^a = \xi^a$ on $X=T$. This rescales $H_{\rm sur}$ but note that the corresponding Rindler temperature is now just $(1/2\pi)$. The simultaneous rescaling  of $H_{\rm sur}$ and $\kappa$ keeps the entropy $S= 2\pi H_{\rm sur}/\kappa = A_\perp/4$ invariant. 

Based on these results, we can give two prescriptions for computing the entropy. The first one is from the surface term in the action: 
\begin{equation}
 S = A_{\rm sur}\Big|_{t=2\pi/\kappa}
\label{time}
\end{equation} 
 which of course has a natural interpretation in the Euclidean Rindler sector since the periodicity of Euclidean time is $2\pi/\kappa$; but as a prescription it can be used even with Lorentzian Rindler time.
The second prescription is essentially that of Wald entropy \cite{wald}:
\begin{equation}
 S = \frac{2\pi}{\kappa}\, \int d^{D-2}\Sigma_{ab}\; J^{ab}
\end{equation} 
In terms of $H_{\rm sur}$ the two definitions given above are equivalent to
\begin{equation}
 S = \int_0^{2\pi/\kappa} dt\, H_{\rm sur}  = \frac{2\pi}{\kappa} \, H_{\rm sur} 
\end{equation} 
This shows that, in either way, the central quantity which captures the thermality of the null surface is its heat content, $H_{\rm sur}=(\kappa/2\pi)S$.

From the heat content $H_{sur}$ of the horizon we can obtain the  heat energy per unit area of the horizon, $H_{sur}/A_\perp=\kappa/8\pi=P$, which also appears as the pressure term in the Navier-Stokes equation
 obtained by projecting Einstein's equation on to the null surface \cite{NS1,NS2} and leads to the equation of state $PA_\perp=TS$. This
 heat energy density on the stretched horizon at   $x^1=const$  with $n_c=\delta_c^1$ is
\begin{equation}
 \mathcal{H}=\frac{NK}{8\pi}
 =\frac{1}{16\pi}\sqrt{-g}V^cn_c=-\frac{1}{16\pi}\,n_c(f^{ab}N^c_{ab})
\end{equation} 
where $f^{ab}=\sqrt{-g} \, g^{ab}$ and $N^c_{ab} $ is the corresponding conjugate momentum defined by
\begin{equation}
N^i_{jk}\equiv\frac{\partial (\sqrt{-g}L_{bulk})}{\partial(\partial_if^{jk})}
=-[\Gamma^i_{jk}-\frac{1}{2}(\delta^i_j\Gamma^a_{ka}+ \delta^i_k\Gamma^a_{ja})]. 
\label{defN}
\end{equation}
Closing the loop, one can express the surface term in the Hilbert action in terms of the heat energy density of the horizon as:
\begin{equation}
A_{sur}=\frac{1}{16\pi}\int dt\int d^2x\ n_c(f^{ab}N^c_{ab})= - \int dt\int d^2x \mathcal{H}
\label{asurN}
\end{equation} 
These results show that  the heat content is directly related to gravitational degrees of freedom $f^{ab}$ and their conjugate momenta $N^c_{ab}$. When expressed in terms of $N^c_{ab}$ the expression is not generally covariant; the $N^a_{bc}$ vanishes in the inertial coordinates but not in the Rindler coordinates.

We conclude this section with a brief mention of the physical importance of $H_{sur}$, which also
arises in the study of black hole horizons \cite{surH} and is closely related to the phase of the semiclassical wave function of the black hole. 
The phase of the semiclassical wave function describing a geometry with a horizon will pick up a term from the surface term in the action given by:
\begin{equation}
 \Psi \propto \exp(iA_{\rm sur}) = \exp\left(-i\int dt H_{\rm sur}\right)
\end{equation} 
When a semiclassical black hole interacts with  matter fields, the probability for its area to change by $\Delta A_\perp$ is  determined  by the integral of the form
\begin{equation}
\mathcal{P}(\Delta A_\perp)=\int_{-\infty}^{\infty} dt F_{m}(t)\exp[-it\Delta H_{sur}]
=\int_{-\infty}^{\infty} dt F_{m}(t)\exp[-it\frac{\kappa}{8\pi} \Delta A_\perp]
\end{equation} 
where $F_m(t)$ is a  suitable matter variable. The exponential redshift near the horizon will result in  the time evolution of $F_m(t)$ having the asymptotic form $\exp[-iC\exp(-\kappa t)]$ where $C$ is a constant. An elementary calculation now shows that   the relative probability for the horizon to change its area by $\Delta A_\perp$ is given by  $\exp[\Delta A_\perp/4]$.

\section{\label{infinitesimal}Infinitesimal coordinate transformation from the inertial to the Rindler frame}

The calculation of $H_{\rm sur}=TS$ in the Rindler frame, performed in the last section, used the full non-linear transformation between the inertial and Rindler coordinates in flat spacetime (or between the freely falling and static coordinate systems near a horizon in a curved spacetime). The calculation based on the Noether potential also used the Killing vector field associated with the (local) Rindler frame. These results bring to focus the issue we raised in Sec. 1 viz., how certain degrees of freedom contributing to the heat content of a null surface arises when we switch from one coordinate system to another. Because degrees of freedom are closely associated with the symmetries of the theories --- which in this case are infinitesimal coordinate transformations --- it would be nice to see whether the results in Sec. 2 can be obtained using only \textit{infinitesimal} coordinate transformations.  As a preliminary to this investigation in Sec.4, we will now set up  the necessary transformation. 

Let us begin with a flat spacetime described in the inertial coordinates ($T,X,\mathbf{X}_\perp$) with the metric
\begin{equation}
ds^2 = -dT^2 + dX^2 + dL_{\perp}^2~,
\label{1.01}
\end{equation}
where $dL_{\perp}^2$ denotes the flat transverse metric which will not play any significant role in the discussion.  The usual coordinate transformation from the inertial coordinates ($T,X,\mathbf{X}_\perp$) to the Rindler coordinates $(t,x,\mathbf{X}_\perp)$  can be taken to be 
\begin{eqnarray}
 T = \kappa^{-1}(1+2\kappa x)^{1/2} \sinh (\kappa t); \quad
X  = \kappa^{-1}(1+2\kappa x)^{1/2} \cosh (\kappa t)-\kappa^{-1}.
\label{1.03}
\end{eqnarray}
leading to the metric:
\begin{eqnarray}
ds^2 = - (1+2\kappa x)dt^2 + \frac{dx^2}{1+2\kappa x} + dL_{\perp}^2~.
\label{1.02}
\end{eqnarray}
These transformations are structured such that they reduce to identity when the acceleration  $\kappa=0$ making the metric in \eq{1.02} reduce to the metric in \eq{1.01}. 

We are interested in the infinitesimal form of these coordinate transformations which we will define as the ones obtained by retaining only terms up to linear order in $\kappa$ in \eq{1.03}. It is easy to verify that, to this order, the transformations in  \eq{1.03} become: 
\begin{equation}
T = t + \kappa xt + {\cal{O}}(\kappa^2); \,\,\,\
X = x +\frac{1}{2}\kappa t^2 -\frac{1}{2}\kappa x^2 + {\cal{O}}(\kappa^2)~.
\label{1.07}
\end{equation}
The inverse transformations are:
\begin{equation}
t = T - \kappa XT + {\cal{O}}(\kappa^2); \,\,\,\
x = X - \frac{1}{2}\kappa T^2  + \frac{1}{2}\kappa X^2 + {\cal{O}}(\kappa^2)~.
\label{1.07n1}
\end{equation}
(Since $\kappa$ is a dimensionfull parameter, the smallness or otherwise of $\kappa$ is somewhat ill-defined. To handle this, we can replace $\kappa$ by $\epsilon \kappa$ in all expressions where $\epsilon$ is a \textit{dimensionless} infinitesimal parameter and do the Taylor series expansion in $\epsilon$. We will not bother to do this since it clutters up the notation and ultimately leads to the same results.)
In general, an infinitesimal transformation between two coordinate systems $x^a \to \bar x^a = x^a + q^a(x)$ is implemented by a vector field $q^a(x)$ in the spacetime. (Once again, $q^a $ should be thought of as $\epsilon q^a$ with all expansions treated as Taylor series in $\epsilon$.) In the case of infinitesimal transformation from the inertial to Rindler coordinate system
the vector field $q^a \equiv x^a-X^a$ has the components in the inertial frame given by:
\begin{eqnarray}
&&q^{T} = t - T = - \kappa XT;
\nonumber
\\
&&q^X = x - X = -\frac{1}{2}\kappa T^2  + \frac{1}{2}\kappa X^2~.
\label{1.08}
\end{eqnarray}
It can be easily verified that $q^a$ is not a Killing vector; its integral curves are given by
\begin{eqnarray}
X^2 T- \frac{T^3}{3}= C~,
\label{1.16}
\end{eqnarray}
where $C$ is a constant (In fact, as far as we know, this vector field has not been studied in the literature).

To avoid possible misunderstanding, we emphasize again how the vector field $q^a$ is related to the transformation from inertial to Rindler coordinate system. This is important because, this transformation in \eq{1.03} does \textit{not} reduce to identity at infinity; similarly, the Rindler metric in \eq{1.02} is \textit{not} asymptotically flat. So one cannot obtain an infinitesimal version of the transformations at asymptotic infinity. Instead, we imagine a \textit{family} of Rindler transformations, parametrized by a dimensionless parameter $\epsilon$ with an acceleration $\epsilon\kappa$. When we vary the parameter $\epsilon$ from zero to unity, we move from identity transformation (at $\epsilon=0$) to the final Rindler transformation (at $\epsilon=1$). Now we can consider transformations in this family, close to the identity by doing a Taylor series expansion of \eq{1.03} in the parameter $\epsilon$. To the linear order in $\epsilon$ we will get $x^a=X^a+\epsilon q^a$ which \textit{defines} the vector field $q^a$ we work with. Once this exercise is carried out we can set $\epsilon$ to unity. As mentioned earlier, it is obvious that --- since $\epsilon$ occurs only in the combination $\epsilon\kappa$ --- the same $q^a$ is obtained by simply expanding the transformation in \eq{1.03} in a Taylor series in $\kappa$. This procedure, in fact, is quite general. If we consider any finite coordinate transformation $\bar x^a=f^a(x^i,\lambda)$ where $\lambda$ is a (in general, dimensionfull) parameter defined such that we get identity transformation when $\lambda=0$, then by considering the family of transformations  
$\bar x^a=f^a(x^i,\epsilon\lambda)$ we can smoothly interpolate between identity transformation and finite transformation when $\epsilon$ varies between 0 and 1. Expanding $\bar x^a=f^a(x^i,\epsilon\lambda)$ in a Taylor series in $\epsilon$ we will get $\bar x^a=x^a+\epsilon q^a$ which identifies the vector field $q^a$.

\section{\label{thermality}Thermality from the infinitesimal coordinate transformation}

We will now relate the infinitesimal coordinate transformations between inertial and Rindler coordinates to the thermality of the null surface $X=T$. This is probably the most instructive example because it shows such an effect can arise \textit{even in the flat spacetime}. The idea generalizes in a straight forward manner to all static horizons and we will comment on these generalizations later on.

How does the surface $X=T$ acquire an interpretation in terms of a heat content $H_{\rm sur}$ when we made a coordinate transformation from inertial to Rindler frame?
Of the two methods described in Sec.2 for computing $H_{\rm sur}$, the one based on surface term in the action is conceptually simpler to understand, which we will discuss first. 

We know that neither the bulk nor the surface term of the Einstein-Hilbert action  is generally covariant individually though, of course, $R=0$ in all coordinate systems.  In the inertial coordinates $\Gamma$s vanish making $A_{\rm sur} = 0 = A_{\rm bulk}$ and hence $H_{\rm sur} =0$; so we cannot attribute any heat content to the null surface $T=X$. But since $A_{\rm sur}$ is not generally covariant, it acquires non-zero value under infinitesimal coordinate transformations $x^a \to x^a + q^a(x)$. The connection generated to the lowest order by this transformation is  given by
\begin{equation}
 \Gamma^a_{bc} = -\partial_b \partial_c q^a.
\label{lingamma}
\end{equation}
For the infinitesimal transformation between inertial and Rindler coordinates in \eq{1.08}, this gives the non-zero components (in the Rindler frame) to be
$\Gamma^t_{tx}  = \Gamma^x_{tt} = \Gamma^x_{xx}= \kappa$.
Using this we find that $V^c$ now has the component
$ V^x = -2\kappa$. 
Therefore, the non-zero surface term in the action which is generated is now given by
\begin{equation}
\mathcal{A}_{\rm sur}=- t\left(\frac{\kappa A_\perp}{8\pi}\right)
\label{surfaceH1}
\end{equation} 
leading to the standard results 
\begin{equation}
H_{\rm sur} = - \frac{\partial A_{\rm sur}}{\partial t} =   \left(\frac{\kappa A_\perp}{8\pi}\right)                                                                     \end{equation} 
and, using Eq. (\ref{time}) we get:
\begin{equation}
 S = A_{\rm sur}\Big|_{t=2\pi/\kappa} = \frac{A_\perp}{4}
\end{equation} 
In this approach, it is not surprising that we get $H_{\rm sur} \ne 0$ because $A_{\rm sur}$ is not generally covariant.
The infinitesimal coordinate transformation from inertial to Rindler coordinates generates a non-zero $V^c$ and thus a nonzero contribution to $A_{\rm sur}$ on the horizon leading to  a non-zero $H_{\rm sur}$. (Aside: Note that this corresponds to the situation mentioned earlier, in the para after \eq{nofl}. When we make an infinitesimal coordinate transformation $x^a\to x^a+\epsilon q^a$, the $\Gamma$ that is produced are of $\mathcal{O}(\epsilon)$. But from the structure of the scalar curvature $R\sim \Gamma^2+\partial\Gamma=\mathcal{O}(\epsilon^2)+\mathcal{O}(\epsilon)$ we see that the two terms are of different order. When $R=0$, this will require the two terms to vanish \textit{separately}, which --- in turn --- implies that $L_{sur}\propto \partial_c(\sqrt{-g}V^c)=0$. The integral of  $L_{sur}$, which is a surface term, will also vanish when the contribution from the \textit{entire} surface is taken into account. However, the contribution of this surface term from \textit{just on one surface} need not vanish as long as  $\sqrt{-g}V^c$ itself is not zero  --- but a constant, which is the current situation.)

What \textit{is} surprising is that the \textit{infinitesimal} coordinate transformations captures the exact result! The transformations in \eq{1.07} are obtained by linearising the full transformations in   \eq{1.03} in $\kappa$.
The $A_{\rm sur}$ was computed correct only to linear order in $\kappa$ because we used the linearized expressions in \eq{lingamma} in its calculation. In spite of this, the result in \eq{surfaceH1} matches with the exact result in \eq{surfaceH}. 

A \textit{partial}, mathematical reason for this  result is the following. Since the exact form of  $A_{\rm sur}$ has to be proportional to $t (A_\perp/L_P^2)$, dimensional considerations show that the proportionality constant must have the dimensions of inverse length. The only such constant available to us is $\kappa$ and hence it follows that the expression should have the form $C(t A_\perp \kappa/L_P^2)$ where $C$ is  a dimensionless numerical factor. Therefore, $A_{\rm sur}$ \textit{must be linear} in $\kappa$ and an infinitesimal transformation accurate to linear order in $\kappa$ can reproduce the result except for a numerical factor. 

This argument, however, does not explain why we obtain the correct numerical factor which is not guaranteed by the dimensional analysis. A more physical reasoning will be to attribute this result to the fact that infinitesimal coordinate transformations are all that is required to study the nature of gauge degrees of freedom. As we conjectured in Sec.1, the observer dependent thermality of null surfaces has to do with upgrading of some gauge degrees of freedom to true degrees of freedom. Infinitesimal coordinate transformations should be capable of capturing this effect, which is what has happened.

We will now turn into the second procedure described in Sec.2 for computing $H_{\rm sur}$ using Noether potential. A natural vector field to use in this context is, of course, $q^a \equiv x^a - X^a$, which defines the infinitesimal 
coordinate transformation between the inertial and Rindler coordinates. We will now compute the right hand side of \eq{noetherint}
using $J^{ab}$ defined for $q^a=x^a - X^a$ and \textit{working entirely in inertial coordinates}. The only relevant non-trivial component of $J^{ab}$ is $J^{T X} =  (\kappa/8\pi)T \approx  (\kappa/8\pi)t$ to order $\mathcal{O}(\kappa)$. Using $d\Sigma_{TX} = -\frac{1}{2} \sqrt{\sigma}\, d^2X_\perp$ we find that  the Noether integral now gives  {\footnote{Incidentally, the null horizon surface for the exact transformations in Eq (\ref{1.03}) is at $T=X+1/\kappa$. If we evaluate the charge on this surface, instead of on the surface $X=T$, and express the result in terms of Rindler time $t$, we will get the same result in Eq (\ref{defQ}).}},
\begin{equation}
 Q =\int d^{2}\Sigma_{ab}\; J^{ab}
 =- T\,  \frac{\kappa A_\perp }{8\pi} 
 \approx - t\,  \frac{\kappa A_\perp }{8\pi} 
\label{defQ}
\end{equation} 
This is \textit{not} the $H_{\rm sur}$ computed earlier; and indeed it is \textit{not} expected to be since the vector field we are using now is $q^a=x^a - X^a$ and not $\xi^a$ and the result of the integration, of course, will depend on the vector field. What is curious is that the expression matches with the expression for $A_{\rm sur}$.  This  could be, at first sight,   surprising for two reasons: (i) It is not obvious why the Noether integral for the {\it infinitesimal} coordinate transformation, calculated in the \textit{inertial} frame,  should give $A_{\rm sur}$, computed in the \textit{Rindler} frame.  (ii) We have now provided a generally covariant expression for $A_{\rm sur}$ which --- when treated as the surface term of the action --- is obviously not generally covariant. 

It is possible to understand both these features by probing a little deeper into the structure of the Noether charge and Noether potential. In general relativity, the Noether current for a diffeomorphism generated by a vector field $q^a$  is given by
\begin{equation}
 J^c  = \nabla_l (\nabla^c q^l - \nabla^l q^c) = 2 R^c_m q^m + \mathcal{V}^c; \quad
\mathcal{V}^c \equiv g^{lm}\pounds_q N^c_{lm}
\label{diffone}
\end{equation}
where $N^c_{lm}$ is the canonical momentum defined in \eq{defN}. In flat spacetime, $R^a_b =0$ and we have $J^c =  \mathcal{V}^c$. Note that  even though $N^c_{lm}=0$ in inertial coordinates, its Lie derivative along a vector field:
\begin{equation}
\pounds_q N^c_{lm} = -\partial_l\partial_mq^c + \frac{1}{2}\Big(\delta^c_l\partial_m\partial_iq^i+\delta^c_m\partial_l\partial_iq^i\Big)
\label{LieN}
\end{equation}
is nonzero! To the lowest order we can interpret the Lie derivative in $\pounds_q N^c_{lm}$ as giving the variation $\delta N^c_{lm}$ under the diffeomorphism generated by $q^a$. But since  $N^c_{lm} =0$ in inertial coordinates, we find that $\delta  N^c_{lm} = \pounds_q N^c_{lm}$ is equal to   $N^c_{lm}$ in the the non-inertial coordinates. Thus, in this case, $\mathcal{V}^c = g^{lm}\pounds_q N^c_{lm} = g^{lm} N^c_{lm}$ where $N^c_{lm}$ now denotes the canonical momenta in the noninertial frame generated by the transformation. It then follows that, in our case, 
\begin{equation}
 J^c  =  \mathcal{V}^c=  g^{lm} N^c_{lm} 
\end{equation} 
The integral in \eq{defQ} over Noether potential $J^{ab}$ can be expressed as an integral over Noether current $J^c$ 
using Stokes theorem: 
\begin{equation}
 Q =\int d^{2}\Sigma_{ab}\; J^{ab} = \int d^{3}\Sigma_c\; J^{c} \to  \int d^3 \Sigma_c\, \mathcal{V}^c = A_{\rm sur}
\label{Q}
\end{equation} 
In arriving at the third equality we have restricted ourselves to the contribution from the horizon surface, as mentioned earlier; in obtaining the last equality, we have used \eq{asurN} giving $A_{\rm sur}$ in terms of $N^c_{lm}$. This explains why in this case the Noether integral leads to $A_{\rm sur}$.

As an aside, we note the following: The dimensions of $A_{\rm sur}$ and $H_{\rm sur}$ are different and it may be surprising at first sight that the integral over $d\Sigma_{ab} J^{ab}$ leads to $H_{\rm sur} $ in \eq{noetherint} and $A_{\rm sur}$ in \eq{defQ}. This arises because $\xi^a$ used to define $J^{ab}$ appearing in \eq{noetherint} is dimensionless while $q^a$ used to define  $J^{ab}$ appearing in \eq{defQ} has the dimensions of length. Any  deformation vector  field, appearing in $x^a \to x^a + q^a$ with length dimensions, cannot lead to a result with dimensions of $H_{\rm sur}$ but will lead to a result with dimensions of $A_{\rm sur}$ in the Noether integral. So the dimensions are consistent.

We can now define the heat content of the null surface through the relation: 
\begin{equation}
 H_{\rm sur}  = -\frac{\partial}{\partial t} \,  \int d^{2}\Sigma_{ab}\; J^{ab}
= - \xi^a \nabla_a\int d^{2}\Sigma_{ab}\; J^{ab}
\label{Hint}
\end{equation}
which shows that the heat content can be given a generally covariant definition even with our $q^a$. 
In the Rindler coordinates we can also write 
\begin{equation}
 \int dt \, H_{\rm sur}  = -\int d^{2}\Sigma_{ab}\; J^{ab};
\label{Hint1}
\end{equation} 
Since the right hand side of the first relation is generally covariant, we expect the integral of $H_{\rm sur}dt$ also to be generally covariant. In the static situation we are considering, we again get the standard result that $H_{\rm sur} = (\kappa A_\perp/8\pi)$.

It should be stressed that the entire calculation given above was carried out in inertial coordinates. The only hint of the Rindler frame was through the vector field $q^a$ which was defined by the \textit{infinitesimal} coordinate transformation connecting the inertial and Rindler frames. After defining $q^a$ by this procedure, we computed the Noether potential corresponding to this infinitesimal diffeomorphism. This Noether integral, evaluated over the null surface $X=T$ in the \textit{inertial} frame, then reproduces the surface term of the action evaluated using the \textit{exact metric in the Rindler frame}. From this expression, we can obtain the heat content by \eq{Hint} working entirely in the inertial frame.  The fact that the Noether potential for the infinitesimal coordinate transformation to Rindler frame is directly related to the degrees of freedom --- which `come alive' through non-zero $N^a_{bc}$ --- is the physical reason why the procedure works. We believe this is a first step in demonstrating in a concrete setting the relationship between coordinate transformations and observer dependent notion of thermality. 

Before we conclude the section, we clarify certain aspects of the above analysis in order to avoid possible misunderstanding. We begin by noting that the Noether charge
\begin{equation}
Q=\int_{\mathcal{S}} d^2\Sigma_{ab} J^{ab}[q]
\label{Qs}
\end{equation} 
is well-defined for \textit{any} 2D surface $\mathcal{S}$ and \textit{any} diffeomorphism $x^a\to x^a+\epsilon q^a$ which introduces the vector field $q^a$. Normally, in the case static horizons, one takes $\mathcal{S}$ to be the horizon surface (usually treated as a limit of a timelike stretched horizon) and takes $q^a$ to be the timelike Killing vector field; i.e $q^a=\xi^a=(1,0,0,0)$. One then finds that $Q$ can be related to the entropy of the horizon. 

\textit{But one can certainly explore the integral in \eq{Qs} over the horizon surface $\mathcal{S}$ for any diffeomorphism $x^a\to x^a+\epsilon q^a$.} The diffeomorphism can be chosen independently of the surface $\mathcal{S}$. In our approach, we make these choices  as follows:
\begin{enumerate}
\item We choose  $\mathcal{S}$ to be the null surface $X=T$ which acts as the horizon for a class of Rindler observers with acceleration $\kappa$. This allows one to define the temperature of the horizon using the standard, finite, coordinate transformations in \eq{1.03}, using, say, periodicity in Euclidean continuation.

\item We obtain $q^a$ from the finite Rindler transformation by a well-defined Taylor series expansion described at the end of Sec.\ref{infinitesimal}. This essentially involves introducing a book-keeping parameter $\epsilon$, defining a family of transformations and picking $q^a$ from the structure of transformations near the identity transformation.

\item The choice of $\mathcal{S}$ and choice of $q^a$ in the above two items are independent of each other. The thermodynamics of $\mathcal{S}$ (like e.g., definition of temperature) arises from the \textit{full} Rindler transformation and could \textit{not} be generated just from $q^a$. This is obvious from the fact that several finite transformations might have the same infinitesimal limit.

\item So, a priori, if we use $\mathcal{S}$ and $q^a$ defined by the above procedure and evaluate the integral in \eq{Qs} we do \textit{not} expect anything physically meaningful to emerge. One would have naively thought that the infinitesimal limit of this transformation does not contain adequate structure to reproduce the temperature etc. originally defined by the full transformation. 

\item The key point of this paper is that, contrary to this naive expectation, the $Q$ in \eq{Qs} evaluated by this procedure captures the thermodynamic features of $\mathcal{S}$, determined originally via the \textit{full} Rindler transformations.
That is, one can indeed reconstruct the
thermality of the horizon by a prescription which  uses \textit{only} the infinitesimal form of the transformation.  If the $\kappa$ of the \textit{infinitesimal} coordinate transformation can be directly related to the temperature associated
with the Rindler horizon at $X=T$, then it is not surprising if we get the horizon thermality  using only the infinitesimal transformation. But this is not the case; one does \textit{not} expect any relation between the parameter $\kappa$ in the
\textit{infinitesimal} coordinate transformation and the temperature associated
with the Rindler horizon at $X=T$ using full Rindler transformation, which makes our result nontrivial and interesting.

\end{enumerate}

We have tried to give detailed reasoning as to why this happens and how it connects up with the broader picture of degrees of freedom etc. We will now study some generalizations of this approach to curved spacetimes.
  
\section{\label{generalised}Generalization to curved spacetime}

The above ideas can be generalized to a wide class of curved, static spacetimes with horizon. One such class is described by metrics of the form 
\begin{eqnarray}
ds^2 = -f(r)dt^2 + \frac{dr^2}{f(r)} + r^2 d\Omega^2~.
\label{3.01}
\end{eqnarray}
with 
$f(r=r_H)=0$ defining the location of the horizon and the surface gravity given by 
$\kappa = f'(r_H)/2 \ne 0$. It is well known that
(see e.g., page 343 of Ref.~\cite{gravitation}) all such metrics allow a transformation to Kruskal-like coordinates with 
\begin{eqnarray}
\kappa T = e^{\kappa r_*} \sinh (\kappa t);\quad
\kappa X = e^{\kappa r_*} \cosh (\kappa t)~,
\label{3.02}
\end{eqnarray}
where the tortoise coordinate $r_*$ is defined by the relation $dr_* = [dr/f(r)]$. Comparison with the transformation between inertial and Rindler coordinates shows that the $T,X$ coordinate system is similar to inertial coordinates and $t,r$ coordinate system is similar to Rindler coordinates. In fact, the $T,X$ coordinate system is a freely falling coordinate system near the horizon and allows us to define locally inertial frames in that region. 

Using the Taylor series expansion in $\kappa$ of the coordinate transformations in \eq{3.02} we can define the infinitesimal version of the transformations and obtain $q^a = x^a  - X^a$.  We then define the Noether potential for this $q^a$ and evaluate the Noether integral as before. The straight forward but somewhat long computation (see Appendix \ref{AppendixA}) now gives the result that
\begin{equation}
  \int d^{2}\Sigma_{ab}\; J^{ab}
 = -t\,  \frac{\kappa A_\perp }{8\pi} 
\label{defQ1}
\end{equation} 
to the lowest order in $\kappa$. 

The rest of the interpretation is similar. The $T,X$ coordinate system and the associated 
freely falling observers near the horizon do not attribute any special properties to the horizon just as the inertial observers do not attribute any thermal properties to the null surface $X=T$. We however know that the static observers outside the horizon attribute a heat content and entropy to the horizon. The above result shows that \textit{infinitesimal} coordinate transformation from the Kruskal-like coordinates to Schwarzschild-like coordinates captures the essence of this phenomena. 

In the case of flat spacetime $N^a_{bc}$ was zero in the inertial frame and non-zero in the Rindler frame. In this case Kruskal-like coordinates we have, in general,  non-zero values for connection at an arbitrary event. However, near the horizon since Kruskal coordinates reduce to a freely falling frame, we have $N^a_{bc} \approx 0$ near the horizon in the $T-X$ plane in Kruskal coordinates. The coordinate transformation to Schwarzschild-like coordinates now generates $N^a_{bc}$ near the horizon, the integral over which leads to the result in \eq{defQ1}. We believe this provides a fairly local description of the heat content of the horizons and an alternate way of interpreting the results \textit{even in} the well known cases like the Schwarzschild metric, de Sitter metric etc..

A natural question to ask would be how general these results are.  We describe below one possible direction of analysis and some preliminary results leaving further details for a future investigation.
Consider a spacetime described by  a metric of the form 
\begin{equation}
ds^2 = -\Omega(u,v,x_{\perp})dudv + dx_{\perp}^2~,
\label{4.01}
\end{equation}
where $u = T-X$, $v=T+X$ are the standard null coordinates. We will consider the spacetime around the null surface $u=0$.
Consider an infinitesimal coordinate transformations: $\bar{u}= u+\xi^u(u,v,x_{\perp})$ and $\bar{v}= v+\xi^u(u,v,x_{\perp})$, such that
\begin{eqnarray}
&&\xi^u(u,v,x_{\perp}) = A_1 + u A_2 + u^2 A_3+\dots;
\nonumber
\\
&&\xi^v(u,v,x_{\perp}) = B_1 + u B_2 + u^2 B_3+\dots~,
\label{4.02}
\end{eqnarray}
where $A_i$ and $B_i$ are functions of $(v,x_{\perp})$ and we have done a Taylor series expansion around $u=0$. If the null congruence associated with  the surfaces $u=$ constant is $l^a$, then the non-affinity parameter is defined by the relation $l^a\nabla_al_b = \lambda l_b$.
This leads to $\lambda(u,v,x_{\perp}) = \partial_v\ln\Omega$ (see Appendix \ref{AppendixB} for details). We can also do a Taylor series expansion of 
$\lambda$ as 
\begin{eqnarray}
\lambda(u,v,x_{\perp}) &=& \lambda_0(v,x_\perp) + uC_1(v,x_\perp)
+ u^2C_2(v,x_\perp)+\dots~,
\label{4.07}
\end{eqnarray}
Repeating the analysis described in the previous sections, one can show that we again get the result:
\begin{equation}
  \int d^{2}\Sigma_{ab}\; J^{ab}
 = - t\,  \frac{\kappa A_\perp }{8\pi} 
\label{defQ2}
\end{equation} 
provided the coefficients of Taylor series satisfies a particular condition (see Eq. (\ref{4.11}) of Appendix \ref{AppendixB}).
Here $\kappa$ is an average over the horizon of the surface gravity given by 
\begin{equation}
\kappa = \int_{u=0} \frac{\sqrt{\sigma}d^2x }{A_\perp} \lambda_0(v,x_\perp)
\label{4.13}
\end{equation}
This shows that our results will hold for a wide class of infinitesimal transformations near a null surface provided the condition in Eq. (\ref{4.11}) of Appendix \ref{AppendixB} holds.  The physical meaning of this condition is unclear and deserves further analysis.

\section{\label{discussion}Discussion}

We provided an explicit example which throws more light on the issues raised in Sec. 1  related to observer dependence of thermal phenomenon. The central idea is: (i) to introduce a freely falling coordinate system ($X^a$) near the horizon and a more standard coordinate system ($x^a$) appropriate to observers who attribute thermality to the horizon; (ii) determine the infinitesimal coordinate transformations connecting these two coordinate systems and thereby determine the vector field $q^a = x^a - X^a$ and (iii) study the integral over a codimension-2 cross-section of the  horizon of the Noether potential $J^{ab}$ appropriate for $q^a$. We found that one can determine the heat content $H_{\rm sur} = TS = E-F$ of the horizon in terms of these quantities and also relate it to the surface term of the gravitational action. The Christoffel symbols vanish in the freely falling frame but are non-zero in the static coordinates natural to the horizon. This difference is closely related to the degrees of freedom contributing to the heat content and entropy of the horizon and remarkably enough, can be captured just from studying the infinitesimal version of the coordinate transformation.

These results are easy to interpret and understand in the case of flat spacetime where the freely falling coordinates are globally defined static coordinates. We showed that the results also generalize, in a straight forward manner, to the standard spherically symmetric static horizons including Schwarzschild, de Sitter spacetimes etc. One possible direction of further work will be to study the infinitesimal coordinate transformation near a general null surface as indicated at the end of last section.

\section*{Acknowledgments}

The research of one of the authors (TP) is partially supported by a J.C. Bose research grant of DST, India.

\begin{appendices}
\section{\label{AppendixA} Static, spherically symmetric spacetimes}
\renewcommand{\theequation}{A.\arabic{equation}}
\setcounter{equation}{0}  
 
We express the Kruskal-like metric using the coordinates ($T,X$) which have dimensions of length as:
\begin{eqnarray}
ds^2 = f(r)e^{-2\kappa r_*} (-dT^2 + dX^2) + r^2 d\Omega^2~,
\label{krusmetric}
\end{eqnarray}
with the transformations:
\begin{equation}
T= \frac{e^{\kappa r_*}}{\kappa} \sinh (\kappa t); \,\,\ 
X + \frac{1}{\kappa} = \frac{e^{\kappa r_*}}{\kappa} \cosh (\kappa t)~.
\label{krustrans}
\end{equation}
Expanding Eq. (\ref{krustrans}) in a Taylor series in $\kappa$ and keeping the terms upto ${\cal{O}}(\kappa)$, we find
\begin{equation}
T = t + \kappa r_* t; \,\,\,\
X = r_* + \frac{1}{2}\kappa t^2 + \frac{1}{2}\kappa r_*^2~.
\label{3.10}
\end{equation} 
The inverse transformations are
\begin{equation}
t = T - \kappa X T; \,\,\,\ 
r_* = X - \frac{1}{2}\kappa T^2 - \frac{1}{2}\kappa X^2~.
\label{3.11}
\end{equation} 
So the components of the generator $q^a \equiv x^a - X^a$ for this transformation are
\begin{equation}
q^{T} = - \kappa X T; \,\,\,\ 
q^X = -\frac{1}{2}\kappa T^2 - \frac{1}{2}\kappa X^2~.
\label{3.12}
\end{equation}
For the metric in (\ref{krusmetric}), the essential component of the  Noether potential, using (\ref{3.12}), is found to be:
\begin{eqnarray}
J^{TX} &=& \frac{1}{16\pi }\Big[\frac{2e^{2\kappa r_*}}{f(r)}\kappa T 
+ \frac{e^{4\kappa r_*}}{f^2(r)}\Big\{\kappa X T \partial_X\Big(f(r)e^{-2\kappa r_*}\Big)
\nonumber
\\
&+& \Big(\frac{1}{2}\kappa T^2 + \frac{1}{2}\kappa X^2\Big)\partial_{T}\Big(f(r)e^{-2\kappa r_*}\Big)\Big\}\Big]~,
\label{3.13}
\end{eqnarray}
and the unit normals are
\begin{equation}
n_a = \Big(0,\sqrt{f(r)}e^{-\kappa r_*},0,0\Big); \,\,\,\ 
m_a = \Big(\sqrt{f(r)}e^{-\kappa r_*},0,0,0\Big)~.
\label{3.14}
\end{equation}
Therefore,
\begin{eqnarray}
d^2\Sigma_{T X} = - d^2x\sqrt{\sigma} \frac{1}{2}f(r)e^{-2\kappa r_*}~,
\label{3.15}
\end{eqnarray}
and so
\begin{eqnarray}
d^2\Sigma_{T X} J^{TX} &=& - d^2x\sqrt{\sigma} \frac{\kappa T }{16\pi }
- \frac{d^2x\sqrt{\sigma}}{32\pi} \frac{e^{2\kappa r_*}}{f(r)}\Big[\kappa X T \partial_X\Big(f(r)e^{-2\kappa r_*}\Big)
\nonumber
\\
&+& \Big(\frac{1}{2}\kappa T^2 + \frac{1}{2}\kappa X^2\Big)\partial_{T}\Big(f(r)e^{-2\kappa r_*}\Big)\Big]~.
\label{3.16}
\end{eqnarray}
Now in the following we shall show that near the horizon ${\cal{H}}$, the second and the last term in Eq. (\ref{3.16}) are of the order $\kappa^2$ and $\kappa^3$, respectively. The second term can be expressed as
\begin{eqnarray}
\frac{e^{2\kappa r_*}}{f(r)}\Big(\kappa X T\Big) \partial_X\Big(f(r)e^{-2\kappa r_*}\Big)
= \Big(\frac{\partial_X f(r)}{f(r)} - 2\kappa\partial_X r_*\Big) \kappa T X~.
\label{3.17}
\end{eqnarray}
Since $X=X(r_*,t)$,
\begin{eqnarray}
\frac{\partial_X f(r)}{f(r)} &=& \frac{1}{f(r)}\Big[\frac{\partial r^*}{\partial X} \frac{\partial}{\partial r_*} + \frac{\partial t}{\partial X} \frac{\partial}{\partial t}\Big] f(r)
\nonumber
\\
&=& \frac{1}{f(r)} \frac{\partial r^*}{\partial X} \frac{\partial f(r)}{\partial r}\frac{\partial r}{\partial r_*}
\nonumber
\\
&=& f'(r)\Big(1-\frac{1}{2}\kappa X + {\cal{O}}(\kappa^2)\Big)~,
\label{3.18}
\end{eqnarray}
where in the last step (\ref{3.11}) has been used. So, near the horizon
\begin{eqnarray}
\frac{\partial_X f(r)}{f(r)} \kappa T X |_{\cal{H}} = {\cal{O}}(\kappa^2)~.
\label{3.19}
\end{eqnarray}
Similarly, $2\kappa (\partial_X r_*) (\kappa T X) = {\cal{O}}(\kappa^2)$. Therefore the second term in (\ref{3.16}), near the horizon, is of the order $\kappa^2$. A similar analysis shows that the third term in Eq. (\ref{3.16}) is of the order $\kappa^3$. So both the terms in (\ref{3.16}) can be ignored giving:
\begin{eqnarray}
d^2\Sigma_{T X} J^{TX}|_{\cal{H}} = - d^2x\sqrt{\sigma} \frac{\kappa T }{16\pi} = - d^2x\sqrt{\sigma} \frac{\kappa t }{16\pi}~,
\label{3.20}
\end{eqnarray}
upto linear order in $\kappa$.
This has been used in computing the integration in Eq. (\ref{defQ1}).

\section{\label{AppendixB}A more general spacetime}
\renewcommand{\theequation}{B.\arabic{equation}}
\setcounter{equation}{0}  
For the metric Eq. (\ref{4.01}) we find that:
\begin{equation}
d^2\Sigma^{uv} = \frac{\sqrt{\sigma}}{\Omega}d^2x
\label{new1}
\end{equation}
and 
\begin{eqnarray}
&&J_{uv}= -\frac{\Omega}{2}\Big[A_2 - \partial_vB_1 + 2uA_3 
 - u\partial_vB_2 + {\cal{O}}(u^2)\Big]
\nonumber
\\
&&-\frac{1}{2}\Big[(A_1+uA_2+{\cal{O}}(u^2))\partial_u\Omega 
- (B_1+uB_2+{\cal{O}}(u^2))\partial_v\Omega\Big]~.
\label{B1}
\end{eqnarray}
Therefore,
\begin{eqnarray}
&&d^2\Sigma^{uv}J_{uv} = -\frac{\sqrt{\sigma}}{2}d^2x\Big[A_2 - \partial_vB_1 + 2uA_3 
- u\partial_vB_2 + {\cal{O}}(u^2)\Big]
\nonumber
\\
&&-\frac{\sqrt{\sigma}}{2}d^2x\Big[(A_1+uA_2+{\cal{O}}(u^2))\partial_u\ln\Omega 
- (B_1+uB_2+{\cal{O}}(u^2))\partial_v\ln\Omega\Big]~.
\label{B2}
\end{eqnarray}

   In the following we will show that $\partial_v\ln\Omega$ is related to the non-affinity parameter corresponding to the null surface $u=$ constant.
Since $u=0$ is the null surface, the normal to it is a null vector. The explicit expression for the null vector $l^a$ in a ($u,v,x_\perp$) coordinate system is:
\begin{eqnarray}
l^a = (0,1,0,0); \,\,\,\ l_a = (-\frac{\Omega}{2},0,0,0)~.
\label{B3}
\end{eqnarray}
The non-affinity parameter $\lambda$ is defined by the relation $l^a\nabla_al_b =  \lambda l_b$. This leads to $\lambda(u,v,x_{\perp}) = \partial_v\ln\Omega$ so that:
\begin{eqnarray}
\ln\Omega = \int \lambda(u,v,x_{\perp})dv~.
\label{B4}
\end{eqnarray}
Expressing since $\lambda(u,v,x_{\perp})$ in terms of the expansion in Eq. (\ref{4.07}) we get:
\begin{eqnarray}
\partial_u\ln\Omega = \int\partial_u \lambda(u,v,x_\perp)dv
= D_1(v,x_\perp) + uD_2(v,x_\perp)+\dots~,
\label{B5}
\end{eqnarray}
where $D_1(v,x_\perp) = \int C_1(v,x_\perp)dv$, etc.
Therefore (\ref{B2}) leads to
\begin{eqnarray}
&&d\Sigma^{uv}J_{uv} = -\frac{1}{2}d^2x\Big[A_2 - \partial_vB_1 + 2uA_3 
- u\partial_vB_2 + {\cal{O}}(u^2)\Big]
\nonumber
\\
&&-\frac{1}{2}d^2x\Big[(A_1+uA_2+{\cal{O}}(u^2))(D_1+uD_2+\dots)
\nonumber
\\
&-& (B_1+uB_2+{\cal{O}}(u^2))\lambda\Big]~.
\label{4.09}
\end{eqnarray}
In the near null surface limit, the above reduces to
\begin{eqnarray}
d^2\Sigma^{uv}J_{uv} &=& -\frac{\sqrt{\sigma}}{2}d^2x\Big[A_2 - \partial_vB_1 
+A_1D_1 - B_1\lambda_0\Big]~.
\label{4.10}
\end{eqnarray}
Now to obtain the required value of the Noether charge given in Eq. (\ref{defQ2}), we need to impose the following condition on the coefficients:
\begin{eqnarray}
\Big[A_2-\partial_vB_1 + A_1D_1 - B_1\lambda_0\Big]_{(u=0)} = v|_{(u=0)}\lambda_0 = 2T\lambda_0\simeq 2t\lambda_0~.
\label{4.11}
\end{eqnarray}
\end{appendices}


\end{document}